\documentstyle[aps,epsf,multicol]{revtex}

\begin{document}

\draft
\title{ X-ray standing wave and reflectometric characterization
of multilayer structures } 
\author{S. K. Ghose and B. N. Dev\cite{email} }
\address{Institute of Physics, Sachivalaya Marg, Bhubaneswar - 751 005, 
India}
\maketitle
\begin{abstract}
Microstructural characterization of synthetic periodic multilayers by x-ray
 standing waves have been presented. It has been shown that the  
analysis of multilayers by combined x-ray reflectometry (XRR) and
 x-ray standing wave (XSW) techniques can overcome the deficiencies 
of the individual techniques in microstructural analysis. While interface
 roughnesses are more accurately determined  by the XRR technique, layer
 composition is more accurately determined by the XSW technique where an 
element is directly identified by its characteristic emission. These
 aspects have been explained with an example 
of a 20 period Pt/C multilayer. The composition of the C-layers due to 
Pt dissolution  in the C-layers, Pt$_{x}$C$_{1-x}$, has been determined by the
XSW technique. In the XSW analysis  when the whole amount of 
Pt present in the C-layers is assumed to be within the broadened interface, it leads 
to larger interface roughness values, 
inconsistent with those determined by the XRR technique. 
Constraining the interface roughness values to those determined by the XRR 
technique, requires an additional amount of dissolved Pt in the C-layers to explain the 
Pt fluorescence yield excited by the standing wave field. 
This analysis provides the average composition Pt$_{x}$C$_{1-x}$ of the C-layers.

\end{abstract}
\begin{multicols}{2}
\narrowtext

\section {Introduction}
	Improvements in the thin film deposition 
techniques in recent years have led to the fabrication of layered synthetic
microstructures (LSM) consisting of 
thin  layers of alternating elements or compounds 
\cite{ref1,ref2}. These materials have unique 
structural\cite{ref3}, magnetic \cite{ref4} and electronic \cite{ref5}
 properties with a wide range of applications. 
 LSM containing alternating layers of high
 atomic number elements (eg., W, Mo, Pt etc.)
and low atomic number elements (eg., C, Si etc.) are being 
used as x-ray reflectors\cite{ref6}. Indeed, x-ray multilayer optics 
are now used in many applications including x-ray astronomy, 
microscopy, spectroscopy, as filters and
monochromators for intense sources such as synchrotron radiation and x-ray 
laser cavities.  It is important to correlate the measured properties with structure so that preparation 
techniques can be optimized to yield high performance materials. 
X-ray techniques are very useful for the measurement
of the microstructural aspects of the multilayered systems. Here we present 
the application of combined x-ray standing wave and x-ray reflectometry
techniques for the microstructural analysis of periodic multilayers.

        For a perfect single crystal, according to the dynamical theory of
x-ray diffraction \cite{ref7,ref8}, a standing wave field is generated
 within the crystal as a result of superposition of the incident
 and the diffracted waves when x-rays are Bragg-reflected by the crystal.
The equi-intensity planes of the standing wave field are parallel
to and have the periodicity of the diffracting planes. At an angle of incidence
corresponding to the rising edge of the diffraction peak the antinodal
planes of the standing wave field lie between the diffracting planes. As
 the angle of incidence increases the antinodal planes move continuously
inward onto the diffracting planes at the falling edge of the
diffraction peak. Over the angular region of Bragg reflection, emission such as
 fluorescent x-rays \cite{ref9,ref10,ref11} and electrons \cite{ref12}
from the crystal is strongly modulated, being maximum (minimum) when the
antinodal (nodal) planes coincide with positions of the atoms in the crystal
or on the surface. By measuring the angular dependence
 of the intensity
of the emitted fluorescence and comparing with the computed angular dependence,
the standing wave field has been used as a structural probe to determine
the positions of the impurity atoms in crystals \cite{ref9,ref10,ref11,ref13},
 adsorbed atoms on surfaces \cite{ref14}, atoms at a layer/substrate
 interface \cite{ref15} and to study thermal effects such as broadening of 
atomic position due to thermal vibration \cite{ref16} and 
order-disorder transitions \cite{ref17}.
Various applications of the x-ray standing wave (XSW) technique to problems 
relating to single crystal surfaces and interfaces may be found 
in recent reviews \cite{ref18,ref19}.

	 The standing wave  phenomenon was also observed in
 multilayer mirrors \cite{ref20,ref21,ref22} and 
Langmuir Blodgett multilayer films \cite{ref23}. 
 This standing wave field was also used in different ways for analyzing
 the local structure of multilayers \cite{ref24,ref25},
 density evaluation of
 deposited films on multilayers \cite{ref26} and 
selective extended x-ray absorption fine structure analysis \cite{ref27}.

	For a periodic multilayer system x-ray reflectivity (XRR) is 
used to determine bilayer periodicity,  interface
roughness and the fractional thickness of the layers in a bilayer.
Interface roughness characterization by x-ray standing waves
has been attempted for a Ni/C multilayer system \cite{ref25}. However,
the extracted parameters  were not optimized. Matsusita et al. \cite{ref28}
used XSW to determine the density of impurity atoms in a multilayer structure.
Here we present a combined reflectivity and standing wave characterization
of a periodic multilayer system to extract various structural parameters.
As an example we use a 20-period Pt/C multilayer system.
Comparing with experimental data we show that, structural parameters
 extracted from x-ray reflectivity 
analysis cannot explain  the Pt fluorescence yield excited by 
x-ray standing waves. Explanation of the Pt fluorescence yield, additionally 
requires the presence of an amount of  dissolved Pt in the C-layers.
XSW  analysis provides the amount of dissolved Pt in C and 
the average composition Pt$_{x}$C$_{1-x}$ of the C-layers. 
Probing a small quantity of material dissolved from one layer into the
other layer of a layer-pair in a multilayer system is very important
for magnetic multilayers where alternating layers are magnetic and nonmagnetic
materials. A small amount (even a few percent) of magnetic impurity (either
from the magnetic layer or external) in the nonmagnetic layer can change 
magnetic coupling and magnetoresistance significantly \cite{ref29}, presumably
because of changes in the topology of the Fermi surface of the resulting
 alloys.. The importance of the combined XSW and XRR analysis is elucidated.

\section{Theory}
We give a brief theoretical
background for the  x-ray standing wave generation inside a multilayer
system.
We mainly follow the treatment given by Dev et al. \cite{ref30}
 for the formation of standing waves and resonance enhancement 
of x-rays in layered 
materials using the recursion method of Parratt \cite{ref31}. 
Then we obtain the field intensity for a periodic multilayer system
and compute the angular variation of fluorescence yield from constituent
elements in the multilayers. The fluorescence yield profile depends 
on the structural parameters of the multilayer. A consistent set of 
microstructural parameters of the multilayer is obtained from the combined
 analysis of reflectivity and fluorescence yield.

\subsection{Reflection from a multilayer system}
\begin{figure}
\protect\centerline{\epsfxsize=2.4in \epsfbox{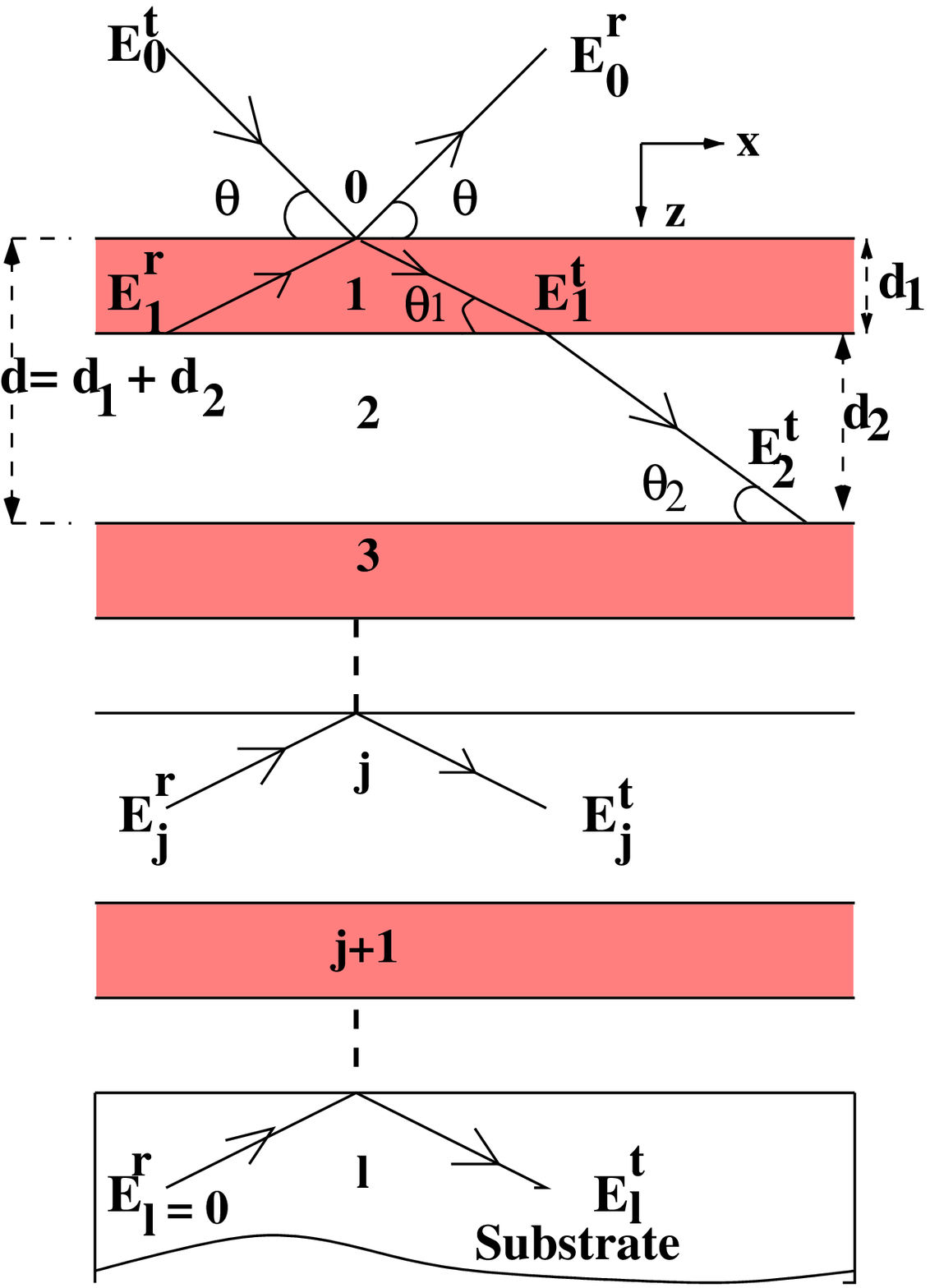}}
\caption{ A schematic representation of x-ray reflection from a 
multilayer system. See text for details. }
\label{fig1}
\end{figure}

If all interfaces are parallel in a multilayer system Fig. \ref{fig1},
a plane electromagnetic wave of frequency $\omega$ 
in a medium {\it j} at a position {\bf r} can be
written as 
\begin{equation}
 E_{j}({\bf r}) = 
E_{j}(0)
exp[i(\omega t - {\bf k}_{j}.{\bf r})]
\end{equation}
where $E_{j}(0)$ is the field amplitude at the top of the ${\it j}$-th layer.

 For all {\it j}, the components of the wave vector,
${\bf k}_{\it j} = {\bf k}_{\it j}^{\prime} - {\it i}
{\bf k}_{\it j}^{\prime\prime}$, are given by
\begin{equation}
k_{j,x} = {2\pi \over \lambda} cos\theta\hskip .1cm ;\hskip .1cm 
{k_{j,z}} = {2\pi \over \lambda}({\epsilon}_{j} - cos{^2}{\theta} )^{1/2}   
\end{equation}
where
${\theta}$
is the glancing angle of incidence, ${\lambda}$ is the wavelength 
of the incident x-rays and the dielectric function $\epsilon_j$ is given by 
\begin{equation}
 {\epsilon}_{j} = 1 - 2{\delta}_{j} - i2{\beta}_{j}
\end{equation}
where
\begin{eqnarray}
{\delta} & = &  {r_e\lambda^{2}\over{2\pi}}N_{0}{\rho_{m}\over M}(Z+f^{\prime})=({\lambda}^2/{2\pi})r_e\rho \nonumber\\
{\beta} & = & {r_e\lambda^{2}\over {2\pi}}N_{0}{\rho_{m}\over M}f^{\prime\prime}=
(\lambda/4\pi)\mu
\end{eqnarray}
In Eqn.(4) N$_{0}$ is Avogadro's number, $\rho_{m}$ is the mass density 
of the element in the layer with atomic number Z and atomic weight M.
$f^{\prime}$ and $f^{\prime\prime}$ are the 
real (dispersive) and the imaginary (absorption) anomalous dispersion
factors, respectively.
$\rho$ is the electron density (including dispersion) and $\mu$ is the linear absorption 
coefficient for the incident photons in the medium.
$r_e$ is the classical 
electron radius.
We consider the medium for the incident
beam to be vacuum with ${\epsilon}_{0}$ = 1.

   For the s-polarization of the electric field and smooth interfaces 
the complex coefficient of reflection $r_{j}$ and transmission $t_{j}$,
 being the ratio of electric fields at the {\it j}, {\it j}+1 
interface, are given by 
Fresnel's formulas
\begin{equation}
r_{j} = {k_{j,z} - k_{j+1,z}\over    
k_{j,z} + k_{j+1,z}}   
\end{equation}
\begin{equation}
t_{j} = {2k_{j,z}\over   
k_{j,z} + k_{j+1,z}}   
\end{equation}
For small $\delta_{j}, \beta_{j}$ approximation, no distinction need be made 
between s-polarization and p-polarization \cite{ref31}. 
   
	For rough surfaces these expressions are to be modified. There are
several methods for obtaining modified expressions. In a 
well-known method \cite{ref32,ref33}
$r_{j}$ is multiplied by a factor $S_{j}$ given by
\begin{equation}
S_{j} = {exp[-2\sigma _{j}^{2}k_{j,z}k_{j+1,z}]}   
\end{equation}
where $\sigma_{j}$ is the root-mean-square deviation of the interface atoms 
from the perfectly smooth condition. An expression like Eqn.(7) is only 
valid for small roughnesses $(\sigma_{j}|k_{j,z}| < 1)$.

   For the modification of $t_{j}$, it is to be multiplied by
\begin{equation}
T_{j} = exp[\sigma_{j}^{2}{(k_{j,z} - k_{j+1,z})}^{2}/2]   
\end{equation}
 
	  So far we have discussed the reflection and refraction at a single
interface. For a multilayer system, involving multiple interfaces, 
the electric fields at all the interfaces can be obtained from either
a recursion relation or from a matrix formalism. In the following we 
will use the method of recursion relation.   
In the recursion method \cite {ref31,ref34}, 
the transmitted field $E_{j}^{t}$ and the 
reflected field $E_{j}^{r}$ at the top of the {\it j}-th layer are found 
from the relations $:$ 
\begin{equation}
E_{j}^{r} = a_{j}^{2}X_{j}E_{j}^{t},   
\end{equation}
\begin{equation}
   E_{j+1}^{t} = {a_{j}E_{j}^{t}t_{j}T_{j}\over    
   1 + a_{j+1}^{2}X_{j+1}r_{j}S_{j}}   
\end{equation}
and       
\begin{equation}
   X_{j} = {(r_{j}S_{j} + a_{j+1}^{2}X_{j+1})\over   
   1 + a_{j+1}^{2}X_{j+1}r_{j}S_{j}}   
\end{equation}
where      
\begin{equation}
   a_{j} = exp(-ik_{j,z}d_{j})   
\end{equation}
$d_{j}$ being the thickness of the {\it j}-th layer.
For the substrate $E_{l}^{r} = X_{l} = 0$ .
  
   The electric field amplitudes $E_{\it j}^{t}$ (transmitted) and 
$E_{\it j}^{r}$
(reflected) can be computed from the knowledge of $\lambda$, $\theta$,
${\epsilon_{\it j}}$'s, the thickness of the layers ($d_{\it j}$'s) and the 
interface roughness ($\sigma_{\it j}$'s)  
using Eqn.(2) through Eqn.(12) and
the reflectivity $\sl {R}$ is then obtained from the ratio of 
E-fields outside the surface :
\begin{equation}
\sl{R(\theta)} = |E_{0}^{r}/E_{0}^{t}|^2 
\end{equation}
For reflectivity from a periodic synthetic multilayer system involving 
interface roughness, this treatment is essentially equivalent to that of 
Underwood and Barbee \cite{ref35}.

	For a periodic multilayer system, below the critical angle of 
incidence, $\theta_{1}^{c}$=$\sqrt{2\delta_{1}}$,
there exists an evanescent wave below the surface and total external reflection
of the incident beam occurs ($|E_{o}^{r}|\approx|E_{o}^{t}|$).
 The interference between $E_{o}^{r}$ and $E_{o}^{t}$ can 
form standing waves above the 
surface \cite{ref22}. For $\theta > \theta_{1}^{c}$, the incident beam penetrates 
into the first layer of the multilayer system.
When $\theta_{1}^{c}\ge\theta_{2}^{c}$, the incident beam penetrates into the 
multilayer system for $\theta > \theta_{1}^{c}$.
If $\theta_{2}^{c} \ge\theta_{1}^{c}$,
there is the possibility of resonance enhancement of x-rays in medium '1'
for $\theta_{1}^{c} < \theta < \theta_{2}^{c}$ \cite{ref30,ref34,ref36}. For $\theta$ greater than both
$\theta_{1}^{c}$ and $\theta_{2}^{c}$, the x-ray beam penetrates into
the multilayer and if the multilayer is periodic, 
Bragg diffractions can occur \cite{ref35} .

	For a periodic multilayer system of x-ray reflectors the
  multilayer period consists of one low and one high electron density 
alternating layers (say, Pt/C/Pt/C ...), the higher density layer works as a marker and the low 
density layer works as a spacer. This arrangement makes the 
system an artificial 
periodic structure. Therefore, in the reflectivity from such a periodic
 multilayer system, Bragg peaks appear at positions determined by
 Bragg's law (including refraction and absorption).
\begin{equation}
2({\it d}_{1}{\it k}^{\prime}_{1,z}+{\it d}_{2}{\it k}^{\prime}_{2,z})=2{\it n}\pi
\end{equation}
or
\begin{equation}
2({\it d}_{1}sin{\theta_{1}}+{\it d}_{2}sin{\theta_{2}})={\it n}\lambda
\end{equation}
where the period of the multilayer is ${\it d=d_{1}+d_{2}}$ and n is the order
of reflection.

	It is well known from the dynamical theory of x-ray
 diffraction from perfect crystals that \cite{ref8}
a  standing wave field is set up in the crystal during diffraction.
 The antinode  position of this wave changes over  half the unit-cell distance in passing the
 diffraction peak. This is also true for x-ray diffraction from 
a periodic multilayer system 
which will be illustrated later. 

\subsection {Field Intensity}
The interference between the incident E-field $(E^{t}_{j})$ and the
reflected E-field $(E^{r}_{j})$ can form standing waves within any layer. 
In order to obtain this standing 
wave field in the $j$-th layer 
one needs to know the fields $E^{t}_{j}$ and $E^{r}_{j}$ 
as a function of depth ($z$). The total E-field at a point 
${\bf r}$ 
in the {\it j}-th layer is given by
\begin{equation}
{\bf{\it E}}_{j}^{T}
({\bf r})
 = {\bf{\it E}}_{j}^{t}
({\bf r}) +
{\bf{\it E}}_{j}^{r}
 ({\bf r})
\end{equation}
where   
\begin{equation}
{ E_{j}^{t}({\bf r})} = 
E^{t}_{j}(0)
exp(-ik_{j,z}z) 
exp[i(\omega t - k_{j,x}x)],
\end{equation}
and   
\begin{equation}
 { E_{j}^{r}({\bf r})} = 
E^{r}_{j}(0)
exp(+ik_{j,z}z) 
exp[i(\omega t -  k_{j,x}x)]
\end{equation}

   Here the origin has been chosen to be on the interface at the top of the  
${\it j}$-th layer. Thus $E_{\it j}^{t}(0)$ and $E_{\it j}^{r}(0)$ represent
the transmitted and the reflected E-fields at the top of the ${\it j}$-th 
layer. 
$E_{\it j}^{t}(0)$ and $E_{\it j}^{r}(0)$ 
are readily obtained from the recursion 
relations (Eqn.(9) through Eqn.(12)). 
The field intensity $I(\theta ,z) 
 = |{\bf{\it E}}_{j}^{T}
({\bf r})|^{2}$
is given by \cite{ref30}
\begin{eqnarray}
& &I(\theta ,z) 
= |{E^{t}_{j}(0)}|^{2}
[exp \{ - 2 k_{j,z}^{\prime\prime} z \} 
 + |{E^{r}_{j}(0)\over
E^{t}_{j}(0)}|^{2} 
exp \{2  k_{j,z}^{\prime\prime} z \}\nonumber \\ 
& & + 2|{E^{r}_{j}(0)\over
 E^{t}_{j}(0)}|cos\{\nu(\theta) + 2k_{j,z}^{\prime}z\}],
\end{eqnarray} 
where $\nu(\theta)$ is defined by  
${E^{r}_{j}(0) \over
E^{t}_{j}(0)} =
{|{E^{r}_{j}(0) \over
E^{t}_{j}(0)} |}e^{i\nu(\theta)}$, i.e., $\nu(\theta)$ is the phase 
of the E-field ratio at the top of the {\it j}-th layer. 
If the absorption in the medium is ignored ( i.e, 
$k_{j,z}^{\prime\prime}$ = 0 ), Eqn (19) reduces to
\begin{eqnarray}
& & I(\theta ,z) 
= |{E^{t}_{j}(0)}|^{2}
[ 1 + |{E^{r}_{j}(0)\over E^{t}_{j}(0)}|^{2} \nonumber \\
& & + 2|{E^{r}_{j}(0)\over
 E^{t}_{j}(0)}|cos\{\nu(\theta) + 2k_{j,z}^{\prime}z\}],
\end{eqnarray} 
It is clear from Eqn. (19) and (20) that a standing wave is
generated within the {\it j}-th layer. 
The quantity within the square bracket in Eqn. (20) may attain a
 maximum value of 4, for
 $|{E_{j}^{r}(0)/E_{j}^{t}(0)|}^{2}$=1.
 For small angles of incidence ($\theta$), in some situations
there are possibilities of resonance enhancement of x-ray intensity 
in the layer. This has been described in details by Dev et al. \cite{ref30}. 
However, at $\theta\gg\theta_{1}^{c}$ and $\theta_{2}^{c}$,  
 $|{E_{j}^{r}(0)/E_{j}^{t}(0)|}^{2} \ll 1$ for a nonperiodic multilayer,
and the field intensity is essentially given by the first term in
 Eq. (19) or (20) with a slight modulation from the second and the 
third terms. For such $\theta$ values the  reflectivity is only 
significant  when $\theta$ satisfies the Bragg condition for reflection
from a periodic multilayer. Standing waves are set up
 in the multilayer when 
Bragg diffraction occurs.
This can be seen from Eqn.(20) by inserting  the Bragg condition [Eq. (14)]
\begin{eqnarray}
& & 2({\it k}_{1,z}^{\prime}{\it d}_{1}+
{\it k}_{2,z}^{\prime}{\it d}_{2})=2{\it k}^{\prime}_{z} d= 2{\it n}\pi\nonumber
\end{eqnarray}
or
\begin{equation}
{\it k}^{\prime}_{z} = {{\it n}\pi\over d}
\end{equation}
where $k^{\prime}_{z}$ is the weighted average value for a layer-pair
of the multilayer with periodicity ${\it d}={\it d}_{1}+{\it d}_{2}$.  
While the magnitude of the E-field ratio varies to some extent
for layer 1 and layer 2 of the bilayer, we can approximate this to be
equal to its value just above the surface, [i.e., $|{E^{r}_{j}(0)\over
E^{t}_{j}(0)}|^{2}\approx ~|{E^{r}_{0}\over E^{t}_{0}}|^{2}$= ~$R(\theta)$,
from Eqn.(13)].
Now for normalized incident intensity, inserting the value
of $k_{z}^{\prime}$ in Eqn.(20) we obtain (for n=1)
\begin{equation}
I(\theta ,z) 
= 1 + R(\theta) + 2\sqrt{R(\theta)} ~cos\{\nu(\theta) +{2\pi\over d}z\}
\end{equation}

It is clear that Eqn.(22) now defines a standing wave
within the multilayer with a periodicity ${\it d}$ and has the same form
as that derived from the dynamical theory of x-ray diffraction from 
 perfect crystals \cite{ref9,ref18}. In the dynamical theory of x-ray
diffraction the E-field in a medium is calculated by solving Maxwell's
equations in that medium and obtaining solutions consistent with the
Bragg's law. This E-field, then, describes the x-ray standing wave
 intensity as a function of angle over the region of the Bragg peak where the phase of ${E^r\over E^t}$($\theta$), $\nu$($\theta$), changes by
 $\pi$ radian \cite{ref8,ref9,ref18,ref37}. The actual value of $\nu$($\theta$)
on the higher-angle side beyond the diffraction peak determines the 
position of the diffraction planes \cite{ref37}. In order to show the similarity between the expressions for the 
standing wave intensity in the dynamical theory for perfect crystals
and in the present case for multilayers we have inserted the Bragg's law
 into Eq. (20) and obtained Eq. (22), which is the well-known form
 obtained from the dynamical theory, where '1/{\it d}' is the magnitude of 
the reciprocal lattice vector for the concerned diffraction. 
The phase variation, $\nu$($\theta$),
for the present case of multilayer is shown in Fig. \ref{fig3}. This has the similar 
form to that obtained from the dynamical theory \cite{ref18,ref37}.

  A periodic multilayer structure can be characterized by generating
 standing waves 
within the multilayer and measuring the 
standing-wave-excited fluorescence yield from one or more elements
present in the multilayer. This is explained in the following sections. For
the computation of standing wave field intensity, $I(\theta,z)$, we will 
use the more rigorous form of Eq.(19).

\subsection {Examples of calculation}

	In this section we  present the results of calculations of 
various quantities in sections II.A and II. B using an example $-$ a 
periodic multilayer system consisting of 20 bilayers of Pt/C 
on a glass substrate. The discussions presented here are general and not
restricted to only Pt/C multilayers.  

 For multilayers, earlier analyses were performed assuming the 
same roughness for both types of
 interfaces (A/B and B/A) in the multilayer (A/B/A/B...) \cite{ref34,ref38}.
 In general, these values should be different. Surface free energy of the 
materials, $\sigma_{A}$ and $\sigma_{B}$, partly control the interface
morphology during the growth. If $\sigma_{A}<\sigma_{B}$, it is the wetting
condition for the growth of material A on material B and a nonwetting 
condition for the growth of material B on material A. Thus A-on-B (A/B)
interface is expected to be smoother. The situation would be reverse for 
$\sigma_{A}>\sigma_{B}$. 
Indeed, high resolution electron
microscopy on W/C multilayers shows that the interface of C growing
 on W is much
sharper than that of W growing on C \cite{ref39}. It must be noted that 
$\sigma_{W}>\sigma_{C}$. However, other factors such as growth temperature
and interdiffusion or chemical reaction between species across the
 interface  also affect the interface roughness \cite{ref40}. 
In any case, there is no 
reason to assume the interface roughness for both types of interfaces to
be equal. Here  we assume different roughnesses for the 
Pt-on-C ($\sigma_{1}$) and the C-on-Pt
($\sigma_{2}$) interfaces. It will be shown later that we indeed get a 
better fit to experimental data when $\sigma_{1}$ and $\sigma_{2}$ 
are allowed to be different. 
 
\begin{figure}
\protect\centerline{\epsfxsize=3.2in \epsfbox{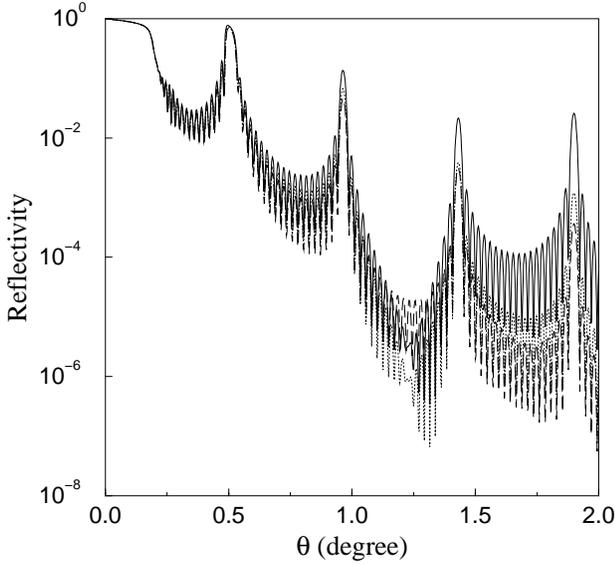}}
\caption{ Reflectivity from a 20 period Pt/C multilayer system with
periodicity $d (43\AA)= d_{1}(17\AA)+d_{2}(26\AA)$ and with surface and 
interface roughnesses ($\AA$) $\sigma_{0}$, $\sigma_{1}$, $\sigma_{2}$
0, 0, 0 3, 3, 3  ($.......$) and 3, 5, 3 
0, 0, 0 (\leaders\hrule width40pt height0.8pt\hskip 30pt ) 3, 3, 3  ($.......$) and 3, 5, 3 
({\bf $-$ $-$ $-$}). }
\label{fig2}
\end{figure}

\begin{figure}
\protect\centerline{\epsfxsize=3.2in \epsfbox{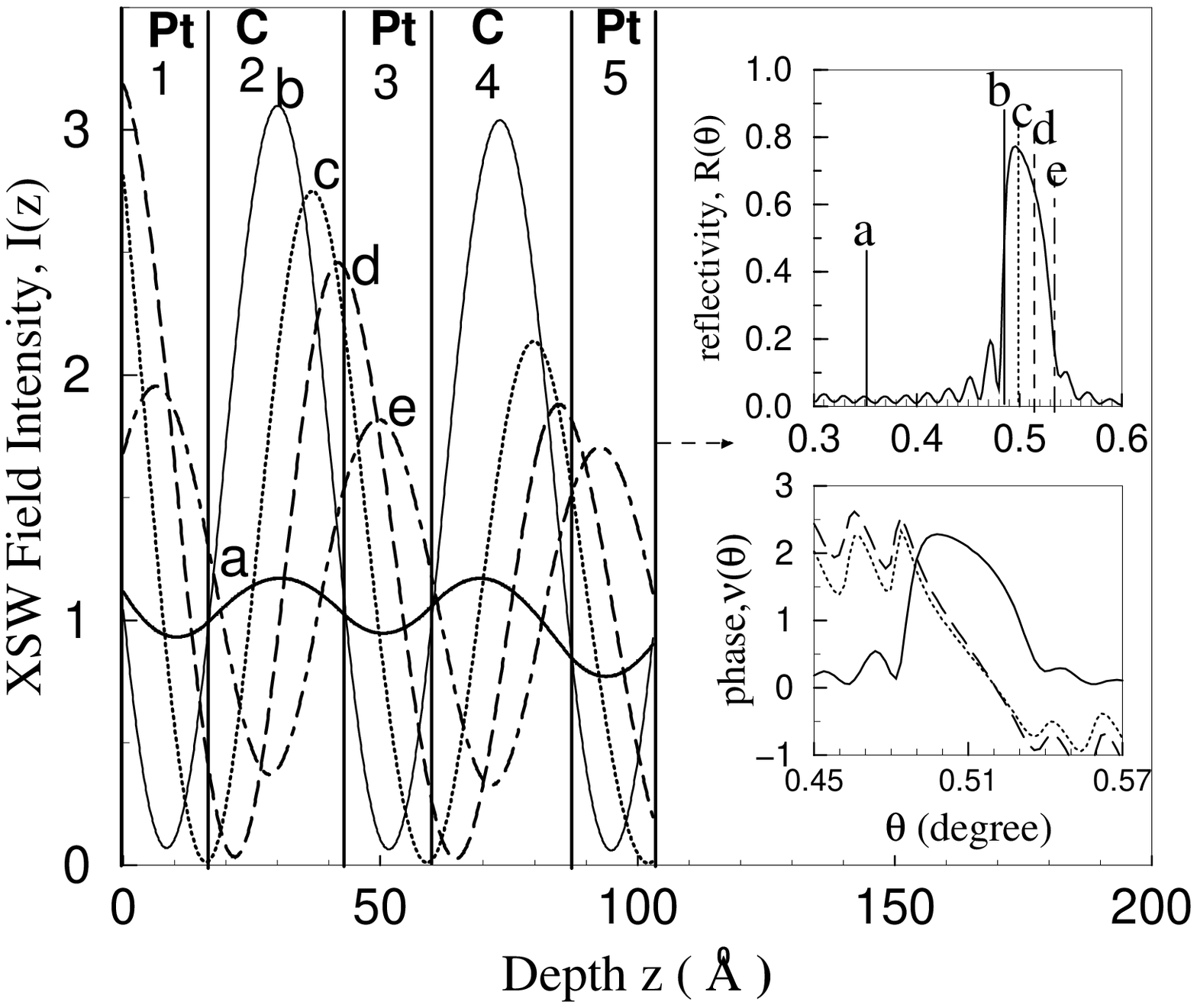}}
\caption{ X-ray standing wave  field intensity distribution within the Pt/C 
multilayer system, at different angles of incidence $\theta$ over the 1st order 
Bragg peak region (shown in the inset).
(a) $\theta$ = 0.35$^{o}$ (\leaders\hrule width40pt height1.8pt\hskip 30pt),
(b) $\theta$ = 0.486$^{o}$ (\leaders\hrule width40pt height0.8pt\hskip 30pt),
(c) $\theta$ = 0.500$^{o}$  (.......), 
(d) $\theta$ = 0.516$^{o}$  ($-$ $-$ $-$),
(e) $\theta$ = 0.535$^{o}$  ($-$ - $-$). The phases, $\nu(\theta)$, 
of E-field ratios ${E_{0}^{r}\over E_{0}^{t}}$ (..........) and
 ${E_{1}^{r}\over E_{1}^{t}}$ ($-$ $-$ $-$) are also shown in the second inset, 
which also shows reflectivity (\leaders\hrule width40pt height0.8pt\hskip 30pt, $\times$3). At a given
 depth {\it z}, the variation in field intensity with angle over the strong
 reflection region occurs mainly because of large variation in phase,
 $\nu(\theta)$. [ see Eqn.(22)].}
\label{fig3}
\end{figure}

\begin{figure}
\protect\centerline{\epsfxsize=3.2in \epsfbox{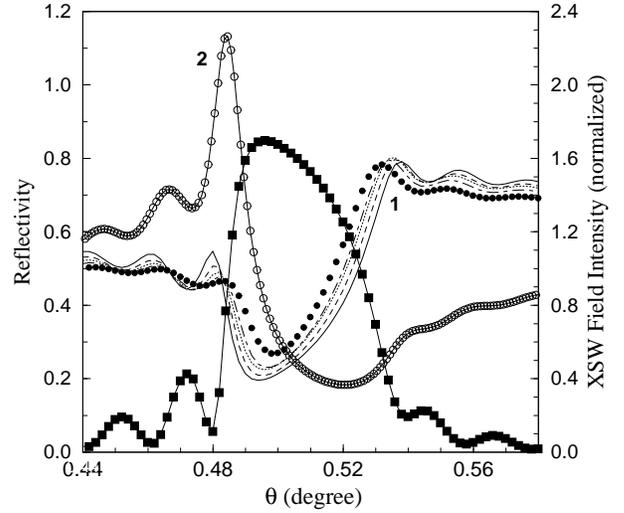}}
\caption{ Integrated XSW field intensity over the Pt-layers and over the C-layers
for different surface and interface roughnesses for the Pt/C multilayer system.
Reflectivity over the 1st order Bragg peak (solid squares) for 
$\sigma_{0}$=0, $\sigma_{1}$=0, $\sigma_{2}$=0 ($\AA$),
integrated field intensity over Pt-layers with
 $\sigma_{0}, \sigma_{1}, \sigma_{2}$ (in $\AA$):
0, 0, 0  (\leaders\hrule width40pt height0.8pt\hskip 30pt); 3, 3, 3   ($-$ $-$ $-$); 
3, 5, 3  (.......); 3, 5, 5  ($-$ - $-$);
3, 7, 7 ($\bullet$ $\bullet$ $\bullet$) and integrated field intensity 
over C-layers (connected open circles)
for $\sigma_{0}$=3, $\sigma_{1}$=5, $\sigma_{2}$=3 ($\AA$).}
\label{fig4}
\end{figure}

	In  Fig. \ref{fig2}., we show the simulated reflectivity curve for smooth
 surface and interfaces alongwith those for several sets of values of surface
 and interface roughness. Total external reflection at low angles and
 multilayer Bragg peaks upto fourth order are seen. The higher
 order peaks are more drastically affected by the surface ($\sigma_{0}$)
and interface roughness ($\sigma_{1}, \sigma_{2}$). The spacing between
 Bragg peaks is determined by the periodicity or the bilayer thickness
($d$). So these parameters can be determined from the reflectivity data
 by a least-squares fitting procedure. In these computations we have used
$\epsilon_{Pt}$= 1 $-$ (2.302$\times$$ 10^{-5}$) $-$ i(2.596$\times$ $10^{-6}$)
and
$\epsilon_{C}$= 1 $-$ (3.016$\times$ $10^{-6}$) $-$ i(8.138$\times$ $10^{-10}$), 
($\rho_{Pt}= 5.05 ~electrons/\AA^{3}$, $\rho_{C}= 0.698 ~electrons/\AA^{3}$ ), 
 $\lambda = 0.709 ~\AA$  ($Mo K_{\alpha_{1}}$ x-rays) and 
$d = 43 ~\AA$ ($d_{1} = 17 ~\AA, d_{2}= 26 ~\AA$).  
X-ray standing wave intensities are shown 
in Fig. \ref{fig3} over the first Bragg peak region ($\theta = 0.3^{o}$ to
$\theta = 0.6 ^{o}$) at several angles shown on the reflectivity curve 
in the inset. The variation of phase, $\nu(\theta)$, of 
${E_{0}^{r}(0)\over E_{0}^{t}(0)}$ and ${E_{1}^{r}(0)\over E_{1}^{t}(0)}$ are shown 
in the second inset of Fig. \ref{fig3}. The field intensity, $I(z)$, can be obtained using
$R(\theta)$ and $\nu(\theta)$ from the insets and Eqn.(22). However,
we have used the more rigorous Eqn.(19) to 
compute the field intensity $I(z)$ at several values of $\theta$.
 At an angle away from the strong
 reflection region (a) the field intensity, $I(z)$, has a weak modulation
 around a value of unity.
At the low-angle side of the diffraction peak (b), there are antinodes of the 
standing wave field in the C-layers (nodes in the Pt layers). As $\theta$ increases 
(b$\rightarrow$c$\rightarrow$d$\rightarrow$e), 
the antinodes shift inward and finally coincide with the Pt-layers.
The field intensity over the Pt-layers gradually increases
 as $\theta$ increases.
The integrated field intensity in the Pt-layers, 
	
\begin{equation}
{\it I_{Pt}}(\theta) =\sum_{j=odd}\int_{0}^{d_{\it j}} I_{j}(\theta , z)dz
\end{equation}
is shown in Fig. \ref{fig4}. $I_{Pt}(\theta)$ for smooth surfaces and interfaces
 ($\sigma_{0}=\sigma_{1}=\sigma_{2}= 0$) and for several sets
 of $\sigma_{0}, \sigma_{1}, \sigma_{2}$ values also shown. It is clearly seen that the field intensity 
$I(\theta)$ variation with $\theta$ is sensitive to surface and 
interface roughness. The integrated field intensity over the carbon layers,

\begin{equation}
{\it I_{C}}(\theta) =\sum_{j=even}\int_{0}^{d_{\it j}} I_{j}(\theta , z)dz
\end{equation}
for $\sigma_{0}=\sigma_{1}=\sigma_{2}=0$ is also shown in Fig. \ref{fig4}. It is 
noticed that the field intensity in the Pt-layers peaks at the high-angle edge
while the intensity in the C-layers peaks at the low-angle edge of the 
reflectivity peak. This opposite trend holds the clue for the determination
 of the concentration of any dissolved Pt in C-layers.

	Our objective is to find the Pt distribution in the Pt/C multilayer.
In the dipole approximation, fluorescence yield from an atom  is proportional 
to the field intensity on the atom. Thus with the measurement of fluorescence
yield from Pt, it is possible to determine the Pt distribution.
Fluorescence yield from Pt in the Pt layers should follow curve '1'
in Fig. \ref{fig4}, while fluorescence yield from Pt in the C-layers should follow
curve '2'. So the effective shape of the fluorescence yield curve will depend
on relative concentrations of Pt in the Pt-layers and the C-layers.

\begin{figure}
\protect\centerline{\epsfxsize=2.4in \epsfbox{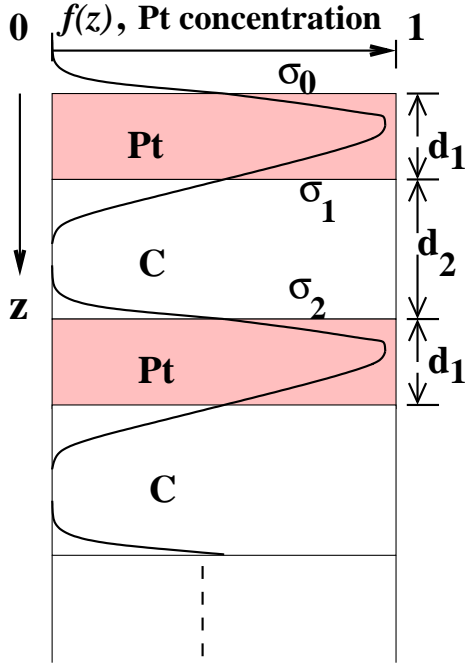}}
\caption{ Schematic diagram of Pt distribution, {\it f(z)}, with
 interface roughness over the bilayer
period. $\sigma_{0}$, $\sigma_{1}$ and $\sigma_{2}$ are 
surface roughness, Pt-on-C and C-on-Pt interface roughnesses, respectively }  
\label{fig5}
\end{figure}

	 Interface roughness can be due to actual roughness or diffusion across the interface.
The Pt distribution, $f(z)$, with interface roughnesses
 $\sigma_{1}\neq\sigma_{2}$ is schematically
shown in Fig. \ref{fig5}. It is obvious that a fraction of Pt is in the C-layers near the interface.
The fluorescence yield of Pt generated ($I^{fg}$) from any depth is proportional to the
product of the field intensity and Pt concentration at that depth.
\begin{equation}
{\it I_{j}^{fg}}(\theta,z)= C {\it I_{j}}(\theta,z){\it f_{j}}(z)
\end{equation}
where {\it C} is a constant.
The fluorescence yield detected outside the sample is given by
\begin{equation}
{\it I_{j}^{fd}}(\theta,z)= C {\it I_{j}}(\theta,z){\it f_{j}}(z)
\times exp[-{\mu_{out}\over {sin\alpha}}(\sum_{m=0}^{j-1}{\it d_{m}}+z)]
\end{equation}
with $d_{0}$=0 and the depth integrated detected fluorescence yield is
\begin{eqnarray}
& & {\it I^{fd}}(\theta)=
 C\sum_{j=1}^{N}{ exp[-{\mu_{out}\over
 sin\alpha}(\sum_{m=0}^{j-1}{\it d_{m}})]}\times \nonumber\\
& & \int_{0}^{\it d_{j}}{\it I_{j}}(\theta,z){\it f_{j}}(z)
exp(-{\mu_{out}\over sin\alpha}z){\it dz}
\end{eqnarray}
where $\alpha$ is the angle between the sample surface and the direction of the 
fluorescence detector from the center of the sample surface, 
and $\mu_{out}$ is the weighted average linear absorption
coefficient for the outgoing (fluorescent) photons.

\begin{figure}
\protect\centerline{\epsfxsize=3.4in \epsfbox{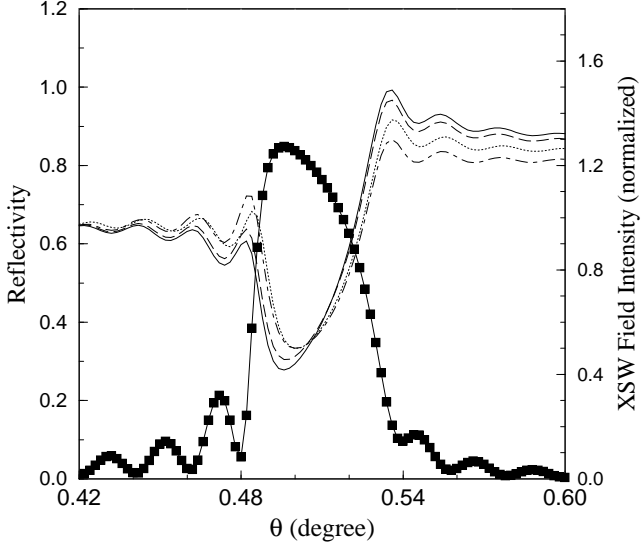}}
\caption{Theoretical plots  for Pt fluorescence yield, computed for 
the distribution of Pt in Fig. \ref{fig5}, over the first order Bragg 
reflection angular region.
Reflectivity (solid squares), Pt fluorescence yield integrated
over Pt-layers with 
surface and interface roughnesses $\sigma_{0}$=3 $\AA$, $\sigma_{1}$=5 $\AA$, 
$\sigma_{2}$=3 $\AA$ (\leaders\hrule width40pt height0.8pt\hskip 30pt), Pt fluorescence yield 
integrated over the whole
 multilayer ($\sigma_{0}$= 3 $\AA$, $\sigma_{1}$= 5 $\AA$, $\sigma_{2}$= 3 $\AA$
and for $f_{c}$=1 ($-$ $-$ $-$), $f_{c}$=0.9 (.......) and
  $f_{c}$=0.8 ($-$ - $-$).  See text for details. }
\label{fig6}
\end{figure}

	The distribution of Pt concentration over the bilayers across the  
Pt-on-C interface is given by
\begin{equation}
{\it f_{1}(z)}= {1\over 2}[1- erf({z\over{\sqrt{2}\sigma_{1}}})]
\end{equation}
for $-d_{1}\leq z\leq d_{2}$, $\it{z}$=0 is on the Pt-on-C interface.
$\sigma_{1}$ is the Pt-on-C interface roughness. Pt distribution
 across the C-on-Pt interface is given by
\begin{equation}
{\it f_{2}(z)}= {1\over 2}[1+ erf({z\over{\sqrt{2}\sigma_{2}}})]
\end{equation}
for $-d_{2}\leq z\leq d_{1}$ where $\it {z}$=0 is taken on the C-on-Pt interface.
 $\sigma_{2}$ is the C-on-Pt interface roughness. The total Pt distribution 
$\it{ f(z)}$ over the bilayer 
and two interfaces is schematically shown in  Fig. \ref{fig5}.
$f(z)=f_{1}(z)+f_{2}(z)$ in the C-layers whereas in the Pt layer $f(z)=f_{1}(z)$ or $f_{2}(z)$, 
whichever is lower. The interface roughnesses
 $\sigma_{1}$ and $\sigma_{2}$ are those used in the analysis of reflectivity.
 Now that the Pt distribution, $f(z)$, over the total
 thickness of multilayer is defined, the integrated detected 
fluorescence yield, $I^{fd}(\theta)$, can be computed using Eqn. (27).
The Pt fluorescence yield 
computed for this distribution of Pt over the first order Bragg reflection
angular region is shown in Fig. \ref{fig6}. 
 
	The solid curve ($\sigma_{0}=3, \sigma_{1}=5, \sigma_{2}=3 ~\AA$)
in Fig. \ref{fig6} shows the computed fluorescence yield profile 
for Pt only in the Pt-layers.
In this calculation the effect of roughness enters only in the
 computation of field intensity
and the contribution to fluorescence yield from  Pt in the C 
layer due to interface broadening
is neglected. This means, in the Eqn.(27), only the sum over
{\it j=odd} layers have been considered. Sum over all layers contain the fluorescence
yield contribution from Pt distributed in the C-layers as well. The 
fluorescence yield curve including this contribution is shown by
the dashed line ($f_{c}$=1, the significance of $f_{c}$
 will be discussed later).

\begin{figure}
\protect\centerline{\epsfxsize=3.4in \epsfbox{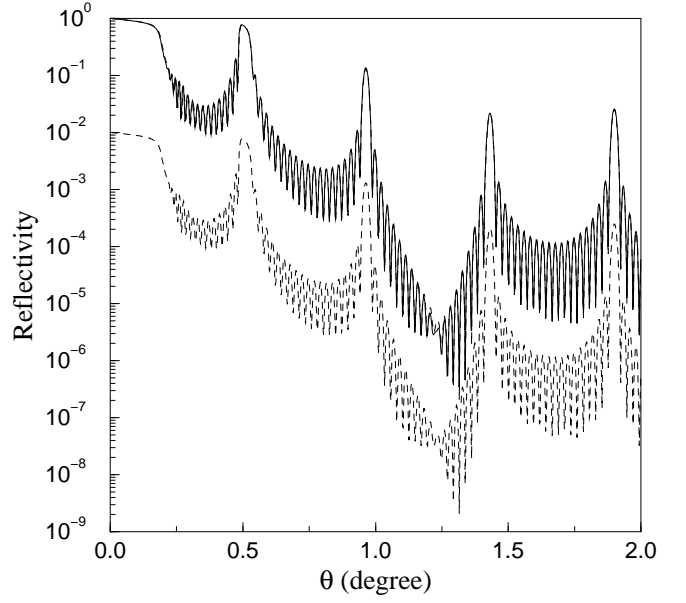}}
\caption{Theoretical plots of reflectivity for different electron densities
$\rho_{C}$ of the C-layers.
$\rho_{C}$ =0.698 $electrons/\AA^{3}$ 
(\leaders\hrule width40pt height0.8pt\hskip 30pt), $\rho_{C}$=0.803 $electrons/\AA^{3}$ ($15\%$ higher
compared to the actual density)
($-$ $-$ $-$). Curves are vertically shifted by two orders. However, they
are also shown in overlaping mode to demonstrate that they are practically 
indistinguishable. }
\label{fig7}
\end{figure}

	The possibility of a small amount of dissolved Pt in the C-layers, 
in addition to the Pt in the interface profile, has not yet been taken into 
account. In the computation of reflectivity the existence of such
 dissolved Pt in C should enter as a change in electron density
of the C-layers. However, due to the low electron density of C 
(0.698 $electrons/\AA^{3}$), reflectivity
is not very sensitive even to a relatively large change in C-layers 
electron density. Reflectivity for a 15\% higher electron density 
(0.803 $electrons/\AA^{3}$) of the C-layers, shown in Fig.\ref{fig7}, is
 hardly distinguishable from that for pure C electron density.
Moreover, the electron density of the C-layers not only 
depends on the amount of dissolved Pt, but also on the change in C-layers 
thickness upon Pt incorporation. The electron density can also change
due to incorporation of ambient atoms (e.g. Ar) during
 multilayer deposition \cite{ref28}.
Thus an accurate determination of the amount of Pt in the C-layers
is difficult from the reflectivity measurement. However, with the x-ray standing
wave method it is possible to determine the amount of dissolved Pt in the
 C-layers through the detection of its fluorescence. Here the detection of Pt is direct and the
 fluorescence yield variation with angle for Pt in the C-layers
has the opposite trend compared to Pt in the Pt-layers (see Fig. \ref{fig4}).
So an analysis of the shape of the measured Pt fluorescence yield curve 
can provide the amount of dissolved Pt in C. 

	We assume  the presence of some dissolved Pt in C.  
Out of total Pt a fraction $f_{c}$ of Pt remains in the Pt-layers
and within the broadened interface regions of the C-layers, and the remaining
fraction ($1-f_{c}$) is dissolved uniformly in the C-layers.
Pt fluorescence yields as a function of angle for $f_{c}$=1, 0.9 and 0.8 are 
shown in Fig. \ref{fig6}. Later we will show with the experimental data 
that the fit to fluorescence yield  
improves when an $f_{c}$$<$1 is allowed in the least-squares fitting procedure.
 From the amount ($1-f_{c}$) of Pt in the C-layers we can obtain the
 average composition (Pt$_{x}$ C$_{1-x}$) of the C-layers.  In the present
 example, fractions $f_c$=1, 0.9 and 0.8 correspond to 0, 4.4 and 
8.7$\%$ Pt  in the C-layers,
respectively. Keeping $f_{c}$=1,
it is also possible to fit the fluorescence data
 assuming broader interfaces, i.e., allowing larger values of $\sigma_{1}$
and $\sigma_{2}$ in Eqns.(28) and (29). However, it would be inconsistent with the
values of $\sigma_{1}$ and $\sigma_{2}$ obtained from the
analysis of reflectivity data, as will be shown in section IV. In order to
 obtain a consistent set of microstructural parameters, it is necessary
to allow, an $f_{c} <$ 1. $f_{c}$ may be called {\it coherent fraction} and 
($1-f_{c}$) {\it incoherent fraction} in analogy with the XSW
 analysis with Bragg 
diffraction from single crystals \cite{ref41}.

\section {Experimental} 

Pt/C periodic multilayers with different bilayer period lengths $d$ ranging
 from 35 $\AA$ to 47 $\AA$ were made on  float glass substrates,
 kept at room temperature, by dc magnetron 
sputtering specially designed for coating inner walls of cyllindrical surfaces.
 Two sputter sources of Pt and C are located at top and bottom of the 
cyllindrical vacuum chamber. Samples were grown at low argon pressure
 of 1 mbar. The deposition rate of Pt and C was 
1 $\AA$/sec and  0.4 $\AA$/sec, respectively.
The layer thickness during deposition was controlled using ion current
 and deposition time. Uniformity in the horizotal plane is achieved by rotating
 the sample while vertical uniformity is acieved by the mask. The overall
 thickness variation was found to be 
$<$2$\%$ over an area of 10 cm$\times$10 cm. The control of the thickness 
of individual layers was within 1 $\AA$. A total of 20 layer
 pairs of Pt/C were deposited in each case. The x-ray specular reflectivity
 measurements have been made on these samples \cite{ref42} to determine
 the bilayer thickness and interface roughness. We have used one of these
 samples for the combined x-ray standing wave and reflectometry analysis.

\begin{figure}
\protect\centerline{\epsfxsize=3.4in \epsfbox{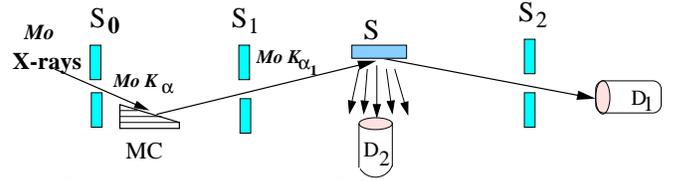}}
\caption{A schematic view of the experimental set up with an asymmetric 
Si(111) crystal monochromator (MC) and incident x-rays from an 18 kW rotating
Mo anode x-ray generator.
Slits: $S_{0}$, $S_{1}$ (horizontal width = 4 mm, vertical width = 100 $\mu$m), 
$S_{2}$ (horizontal width = 10 mm, vertical width = 150 $\mu$m); 
 D1 : NaI(Tl) scintillation detector ; D2 : Si(Li)
 energy dispersive detector; S: sample. }
\label{fig8}
\end{figure}

  Experiments were performed in our laboratory with a 18 kW Mo rotating 
anode x-ray source. The experimental set-up is schematically shown in Fig. \ref{fig8}.
Monochromatic Mo $K_{\alpha_{1}}$ beam  is obtained
with the help of an asymmetrically cut Si(111) crystal monochromator.
The asymmetrically-cut crystal reduces the divergence of the
 monochromatized beam
and is in standard use in X-ray standing wave experiments \cite{ref9}.
 The incident  beam on the sample has an angular divergence of 0.006$^{o}$. 
The vertical beam width is kept as small as 100 $\mu m$.
 Reflected x-rays were detected with a NaI(Tl) detector and the
 Pt $L_{\alpha}$ fluorescent x-rays were detected with a
 Si(Li) detector. The reflected x-rays and  
the fluorescent x-rays were collected simultaneously at each angle.
Control of the instruments for the operation  of the HUBER diffractometer
 and data collection is obtained through a PC using Turbo C programming for
 IEEE and RS-232 protocols. More details about the set-up has
 been presented elsewhere \cite{ref43}.
The average exit angle 
$\alpha$ (the inclination of the Si(Li) detector with respect to the 
sample surface) for fluorescent photons was $50 ^{o}$.
  
\section {Results and discussions}

\begin{figure}
\protect\centerline{\epsfxsize=3.4in \epsfbox{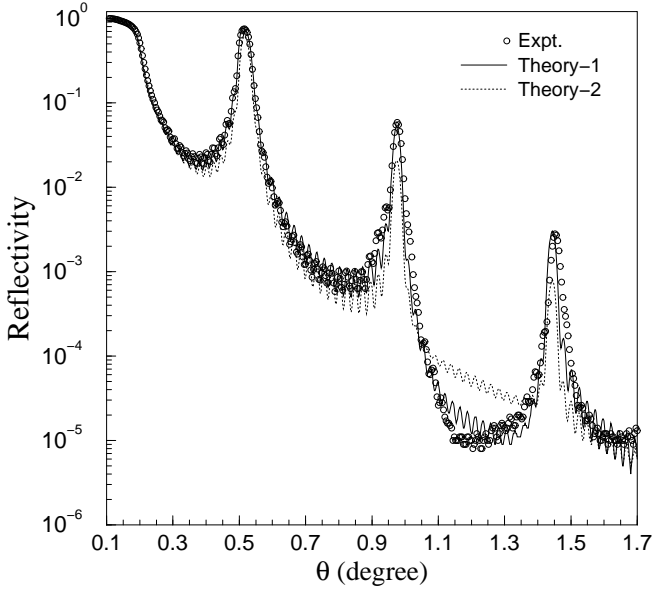}}
\caption{Experimental reflectivity data ($\circ$ $\circ$ $\circ$)
and fitted theoretical reflectivity curve
 (\leaders\hrule width40pt height0.8pt\hskip 30pt)
for a Pt/C multilayer on a glass substrate with 20 bilayers.
 Parameters obtained from the fit: bilayer thickness $d=42.9 ~\AA$, Pt
 layer thickness $d_{1}=16.8 ~\AA$ and C-layers thickness $d_{2}=26.1 ~\AA$, 
surface roughness $\sigma_{0}= 3 ~\AA$, Pt-on-C interface roughness 
$\sigma_{1}=4.5 ~\AA$ and C-on-Pt interface roughness $\sigma_{2}=2.9 ~\AA$.
Theoretical reflectivity curve (..........) for
$\sigma_{0}= 3 ~\AA$, $\sigma_{1}=8.9 ~\AA$ 
and $\sigma_{2}=4.2 ~\AA$ and all other parameters
unchanged. See text for details. }
\label{fig9}
\end{figure}

The experimental reflectivity data and the  
fitted theoretical reflectivity curve (Theory-1) are shown in Fig. \ref{fig9}.
Bragg peaks upto the third order are seen. The small 
oscillations are due to the total thickness of the multilayer. Experimental
data have been fitted by allowing the variation in the electron density,
layer thickness, and surface and interface 
roughnesses  of the layers. From the least-squares fitting the values
of the parameters have been extracted. This fitting gives the Pt-layers density
 $\rho_{1}$=4.95 {\it electrons}/$\AA^{3}$, thickness
 {\it d}$_{1}$= 16.8 $\AA$ and C-layers density (fixed)
$\rho_{2}$=0.698 {\it electron}/$\AA^{3}$, thickness
 {\it d}$_{2}$= 26.1 $\AA$, $\sigma_{1}$=4.5 $\AA$ and $\sigma_{2}$=2.9 $\AA$.
 So the bilayer thickness  
 is 42.9 $\AA$. The third order  peak
 position does not fit properly. This may be due to the  multilayer
having a slight variation in  bilayer thickness along the growth direction.

	 It has been demonstrated that in case of single
 layer films  the roughness is correlated with the thickness of the film
\cite{ref40,ref44}. But in the case of multilayer systems with alternating
 marker and spacer layers the roughness becomes complicated depending
 on the types of material, their diffusion properties, reaction 
and growth behavior \cite{ref40}.
It has been shown that in a W/C multilayer system the W-on-C
 interface is more rough than the C-on-W interface \cite{ref39}.
Fundamentally this is expected because of the nonwetting condition  in
the surface free energy ($\sigma_{W}>\sigma_{C}$) for the 
growth of W on C. In our case $\sigma_{Pt}>\sigma_{C}$ and we also observed 
the same trend: the Pt-on-C interface is more rough
 ($\sigma_{1}$=4.5 $\AA$) than the C-on-Pt interface
($\sigma_{2}$=2.9 $\AA$).
Pt electron density for this sample is 4.95 {\it electron}/$\AA^{3}$, which is
lower than that of pure Pt electron density of
 5.05 {\it electron}/$\AA^{3}$ ($\rho_{m}$=21.5 gm/cc).
In general, thin films tend to have a lower density compared to pure bulk
material. Additionally, interdiffusion across the interfaces leading to a 
mixed layer would decrease the Pt-layers density and
 increase the C-layers density.

\begin{figure}
\protect\centerline{\epsfxsize=3.2in \epsfbox{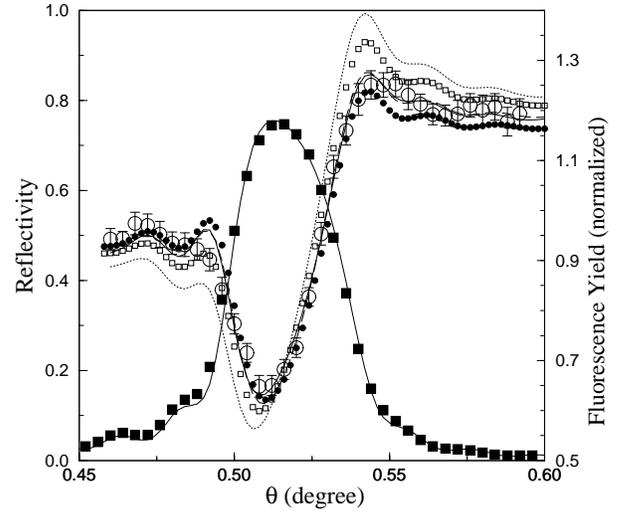}}
\caption{Experimental Pt $L_{\alpha}$ fluorescence yield ( O O O )
and reflectivity (solid squares) vs angle of incidence $\theta$ over the first order Bragg reflection and 
the theoretical curves:
(.......) $\sigma_{0}$=3 $\AA$, $\sigma_{1}$=4.5 $\AA$, 
$\sigma_{2}$= 2.9 $\AA$, $f_{c}$=1.0 (no Pt in C-layers);
 (\leaders\hrule width40pt height0.8pt\hskip 30pt)
 $\sigma_{0}$=3 $\AA$, $\sigma_{1}$=4.5 $\AA$, $\sigma_{2}$=2.9 $\AA$,
 $f_{c}$=0.87 (Pt$_{0.05}$C$_{0.95}$); ($-$ $-$ $-$) $\sigma_{o}$= 3 $\AA$, 
$\sigma_{1}$=8.9 $\AA$,
$\sigma_{2}$=4.2 $\AA$, $f_{c}$=1.0. Also shown, (for $\sigma_o$=3 $\AA$,
$\sigma_1$= 4.5 $\AA$ and $\sigma_2$=2.9 $\AA$) are the curves 
(open squares) for Pt$_{0.03}$C$_{0.97}$ ($f_c$=0.935) and (filled circles) for 
Pt$_{0.07}$C$_{0.93}$ ($f_c$=0.844). Fluorescence curves have been normalized
at $\theta$=0.4$^o$ as in Fig. \ref{fig6}. See text for details. }
\label{fig10}
\end{figure}

	The Pt $L_{\alpha}$ fluorescence yield has been measured
over an angular region containing the first order Bragg peak and 
analyzed as follows.  From the spectrum at each angle in the multichannel
analyzer  only Pt $L_{\alpha}$ peak is selected.
These peaks at all angles are fitted and the background-subtracted 
area is determined. This area gives the yield. This raw yield data 
have been corrected for footprint, probing thickness variation and finite
detector aperture. These corrections are explained at the end of this section.
This corrected Pt $L_{\alpha}$ fluorescence yield vs. angle
 along with reflectivity
over first Bragg peak is shown in Fig. \ref{fig10}. We fit the fluorescence yield
data based on the model described earlier. This model
 incorporates all the parameters extracted from the reflectivity fit.
That means that the density, thickness, surface and interface roughness
 etc.. of the layers are kept intact.
Here we have considered the contribution of roughness
as error functions [Eqns.(28) and (29)] at both the interfaces
with  $\sigma_{1}$=4.5 $\AA$ and $\sigma_{2}$= 2.9 $\AA$.
These $\sigma$-values are the roughness values obtained from the analysis of
reflectivity. (It is well known that reflectivity calculations using explicit
error-function concentration profile at the interface and the flat interface reflection coefficient multiplied by a Debye-Waller function [Eqn. (7)] are
equivalent \cite{ref45}).
If we consider that there is no dissolved Pt in the C-layers (i.e. $f_{c}$= 1),  we do not obtain a good fit.
The best fit is obtained  with the model
with a uniform mixing of Pt in the C-layers with $f_{c}$=0.87.
This means that 13\% of total Pt is dissolved within the C-layers.
Converted to atomic concentration 
this corresponds to the average composition 
Pt$_{0.05}$C$_{0.95}$ of the carbon layer. It should be noted that Pt 
concentration in the C-layers is actually higher near the interface.
 This concentration varies with the distance from the interface and
 can be easily determined from the distributions [Eqns.(28) and (29)].
 	
	In order to show the sensitivity of the fluorescence yield
 curve to the Pt concentration in the C-layers, we also show the plots
for Pt$_{0.03}$C$_{0.97}$ and Pt$_{0.07}$C$_{0.93}$ in Fig. \ref{fig10}. They are
 distinctly different from the data and the fitted curve for 
Pt$_{0.05}$C$_{0.95}$. This clearly shows that the uncertainty in
 the estimated Pt concentration of 5$\%$ is smaller than 2$\%$.
 In the fitting of data the weighted R-factors are 0.041, 0.031, 
0.023, 0.024 and 0.029 for 3, 4, 5, 6 and 7$\%$ Pt, respectively. It is noticed 
from Fig. \ref{fig10} that with increasing Pt concentration in C, Pt fluorescence yield 
increases on the low-angle edge and decreases on the high-angle edge of the
reflectivity curve. This can be easily understood from Fig. \ref{fig3}. At angular
 position 'b' on the reflectivity curve, X-ray intensity is high in the 
C-layers and low in the Pt-layers. However, if there is no Pt in the C-layers,
there would be no Pt fluorescence emission from there. As some Pt migrates from 
the Pt-layers to C-layers, the amount of Pt present in the C-layers would
produce strong fluorescence emission. That is why increasing Pt concentration
in the C-layers produces higher fluorescence yield at this angular position
'b' as seen in Fig. \ref{fig10}. It is also noticed from Fig. \ref{fig3} that the maximum
field intensity in the C-layers is much higher than the maximum field 
intensity in the Pt-layers (see also Fig. \ref{fig4}). This is due to lower absorption
of X-rays in the C-layers. Due to this fact, a given amount of Pt in the 
C-layers produces a stronger fluorescence signal than the same amount in the
 Pt-layers when the X-ray intensities are maximum in the respective layers.
	
	Probing the quantity of material dissolved
from one layer into the other layer of a layer-pair in a multilayer 
system is not only important
for optical mirrors and devices, but also 
very crucial for  magnetic multilayers where interface broadening and 
alloying within the layers affect magnetic properties of multilayers.
In magnetic multilayers with alternating layers of magnetic and non-magnetic
materials, a small amount (even a few percent) of magnetic impurity in the 
nonmagnetic layers can change the magnetic coupling and magnetoresistance.
In fact, in magnetic multilayers with a wide range of 
Cu$_{1-x}$Ni$_{x}$ ({\it x}=0.04 to 0.42) alloy spacer, the smallest 
amount of impurity ({\it x}= 0.04) has shown the largest
change in magnetoresistance \cite{ref29}.
The magnetic impurity in the nonmagnetic layer
of the  multilayer may be an element other than the magnetic element
 present in the multilayer.
Since x-ray fluorescence can identify the element, the distribution of such
impurity elements in the multilayer can be determined by XSW
experiments \cite{ref28}.

     It must be mentioned here that the fluorescence data can also be fitted,
 without assuming the dissolved fraction (i.e. keeping $f_{c}$=1),
 by allowing $\sigma_{1}$ and
$\sigma_{2}$ to vary for the fluorescence fit. This fit is also shown in Fig. \ref{fig10}.
However, the $\sigma$-values obtained from this fit
( $\sigma_{1}=8.9 ~\AA$, $\sigma_{2}=4.2 ~\AA$)
are inconsistent with those obtained from the reflectivity fit.
 The computed reflectivity for these $\sigma$-values,
 as shown in Fig. \ref{fig9} (Theory-2),
is very different from the measured reflectivity. This shows that this set
of larger $\sigma$-values does not represent correct interface roughness.
This is probably the reason why a very
 large $\sigma$-value (10 $\AA$) fitted the fluorescence data of
Kawamura et al. \cite{ref24}.
Our results underline the
necessity for the combined x-ray standing wave and reflectivity analysis
of periodic multilayers. 
We suggest that a combined use of reflectivity and x-ray standing
 waves can provide the microstructural details of a periodic
 multilayer. The procedure to follow is: 
(i) obtain bilayer periodicity, fractional thickness of the high-Z layer
and surface and interface roughnesses from the reflectivity fit,
(ii) interface roughness should not be constrained to be equal
 for both types of interfaces, and 
(iii) use the parameters obtained from the reflectivity fit and proceed for the
the fluorescence data fit with the assumption of a dissolved fraction
of one material in the other, either in uniform distribution
 or with any other improved distribution model. For a more accurate 
determination of this distribution, higher order Fourier components of the 
distribution can be determined by XSW measurements with higher order 
Bragg diffractions.

	In order to fit the reflectivity data to Eqn.(13) and fluorescence data
to Eqn.(27), the following corrections to data were applied: (i) Footprint
correction \cite{ref46} was applied to both reflectivity and fluorescence data. 
At very small angles the beam projection is larger than the sample area. So, 
only a fraction of incident photons are actually incident on the sample. After this 
correction, the data represent what they should be if all the photons
were incident on the sample. (ii) The fluorescence data come from a 
relatively thin layer (thickness of the multilayer) compared to the 
beam penetration depth. Thus with the variation of $\theta$
the effective probe depth changes. To correct for that, 
fluorescence data are to be multiplied by sin$\theta$ at each point.
(iii) The fluorescence detector has a finite aperture and the fluorescent photons may come from a much larger sample area. The detector offers a 
varying effective solid angle for fluorescent photons originating
 from different parts of the sample surface. As the exposed sample
 area varies with $\theta$, this requires a correction which depends
on detector aperture, detector distance from the sample and the sample length.
In our case, over the $\theta$ range (0.45$^o$$-$0.6$^o$) of the first order
Bragg peak region this introduces only a minor correction for 1\% variation
 in detected intensity.	

\section {Conclusions}
	For a periodic multilayer system with alternating layers of a high-Z
and a low-Z  element, Bragg diffraction
of x-rays occurs when the Bragg condition for the bilayer periodicity is
 satisfied. As in diffraction from a large perfect crystal, standing
waves are set up in the multilayer while diffraction occurs. The 
antinodal (or nodal) planes of the standing wave are parallel to the 
layer-planes and have the periodicity equal to the multilayer period.
On the low-angle side of the Bragg-reflection peak, the 
antinodal planes are within the layers with low-Z element. As the angle of 
incidence advances through the diffraction peak the antinodal planes shift
 inward and finally coincide with the nearest layer of high-Z element of the 
layer-pairs. Emission processes, such as photoemission or fluorescence from 
atoms in the multilayer are modulated, over an angular region containing the 
Bragg peak, following the shift of the antinodal
planes. Analysis of this modulation in the emission yield provides structural
information about the multilayer. The usefulness of the combined application
of x-ray reflectivity and x-ray standing wave techniques for the analysis of 
multilayer microstructures has been explained. Deficiencies of each technique
 can be overcome by the combined application of these techniques. XRR depends
on the electron density difference between the layers of the bilayer. Where
electron density of one layer of the layer-pair is very small compared
to the other, reflectivity is not very sensitive to even a large fractional 
change of this electron density. Moreover the change of electron density is
 not necessarily due to the diffusion of atoms from the other
layer of the layer-pair,
it could also be due to other impurities incorporated during multilayer 
fabrication. Thus accurate determination of the layer composition
from XRR technique is practically impossible. These aspects have been 
elucidated with an example of a 20 period Pt/C multilayer. In the XSW 
technique, elements are directly identified. Thus the amount of dissolved Pt
or any other impurity in the C-layers, such as Ar, 
often incorporated during multilayer  
fabrication, can be determined. As interface roughness drastically affects 
the higher order Bragg peaks and  overall intensity at higher angles,
interface roughnesses are more accurately determined by fitting the 
reflectivity data over a large range of angle of incidence. On the 
other hand, in the XSW analysis, if the amount of Pt in the C-layers is assumed
to be solely within the broadened interface and treated as roughness, one 
obtains too large roughness values compared to those obtained from the 
reflctivity fit. Fixing the interface roughness values at 
those obtained from the
 XRR analysis and assuming the remaining Pt to be in uniform
 distribution in the C-layers,
the Pt concentration in the C-layers is determined. (More details about the 
elemental distribution, such as higher order Fourier components, can be 
obtained by XSW measurements with higher order Bragg peaks). Thus a combined
 analysis by XSW and XRR techniques removes the deficiencies of the individual 
techniques. For a 20 period Pt/C multilayer system interface roughnesses
( Pt-on-C: 4.5 $\AA$, C-on-Pt : 2.9 $\AA$) and the C-layers composition 
( Pt$_{0.05}$C$_{0.95}$) have been determined. Determination of a small 
quantity of impurity, even a few percent, in the spacer layer is particularly important 
in the magnetic multilayers.

\acknowledgments


	We thank Dr. G. Lodha and Prof. K. Yamashita for providing 
the Pt/C multilayer sample.


\end{multicols}
\end{document}